\documentstyle[11pt,epsfig]{article}
\title{\bf Accelerated expansion in modified gravity with a Yukawa-like term}
\author{K. Atazadeh and H. R. Sepangi\thanks{email: hr-sepangi@sbu.ac.ir}
\\ {\small Department of Physics, Shahid Beheshti University, Evin,
Tehran 19839, Iran}}
\begin{document}
\maketitle
\begin{abstract}

We discuss the Palatini formulation of modified gravity including
a Yukawa-like term. It is shown that in this formulation, the
Yukawa term offers an explanation for the current exponential
accelerated expansion of the universe and reduces to the standard
Friedmann cosmology in the appropriate limit. We then discuss the
scalar-tensor formulation of the model as a  metric theory and
show that the Yukawa term predicts a power-law acceleration at
late-times. The Newtonian limit of the theory is also discussed in
context of the Palatini formalism.
\end{abstract}
\vspace{2cm}
\section{Introduction}
The expansion of the universe is currently undergoing a period of
acceleration which is directly measured from the light-curves of
several hundred type Ia supernovae \cite{1,2} and independently
from observations of the cosmic microwave background (CMB) by the
WMAP satellite \cite{3} and other CMB experiments \cite{4,5}.
However, the mechanism responsible for this acceleration is not
well understood and many authors introduce a mysterious cosmic
fluid, the so called dark energy, to explain this effect \cite{6}.
Recently, within the framework of higher order gravity theories
\cite{farhoudi}, it has been shown that such an accelerated
expansion could be the result of a modification to the
Einstein-Hilbert action \cite{7}. One such modification has been
proposed in \cite{8} where a term of the form $R^{-1}$ was added
to the usual Einstein-Hilbert action, whose origin can be traced
back to the string/M-theory \cite{nojiri1}. It was then shown that
this term could give rise to accelerating solutions of the field
equations without dark energy.

Based on this modified action, we first use the Palatini
variational formalism to derive our field equations. In the
Palatini formulation, instead of varying the action with respect
to the metric, one views the metric and connection as independent
field variables and vary the action with respect to both
independently. In the original Einstein-Hilbert action, this
approach gives the same field equations as the metric variation.
However when nonlinear terms are added to the action, Palatini
formalism leads to different dynamical equations. The importance
of this formalism lies in the fact that, contrary to the theory
resulting from the metric variation which explains the accelerated
expansion of the universe but is in conflict with solar system
experiments, it is free of such conflicts. Interestingly, the
Palatini formalism has been shown to be equivalent to a
scalar-tensor type theory \cite{12} in which the scalar field
kinetic term is absent from the action \cite{9}. These results are
important and fundamental for this formalism. Also in \cite{14},
Nojiri and Odintsov have shown that a combination of the $R^{-1}$
and $R^{2}$ terms can drive both the current acceleration and
inflation. The Palatini form of this theory is studied in
\cite{11} where the authors obtain the current acceleration of the
cosmic which is compatible with observations. In  \cite{17} the
authors propose a $\ln R$ term to account for the current
acceleration and also inflation. The Palatini formulation of $\ln
R$ gravity is studied in \cite{13} where the acceleration of the
cosmic in the limit of small curvature is predicted. For a more
recent and comprehensive review, the reader should consult
\cite{noj} for models with modified gravity and \cite{gian} for
models with the Palatini approach.

In this paper, we have used the Palatini formalism in a theory
where a Yukawa type term has been added to the action. Since the
term $R^{-1}$, mentioned above, is similar to the Yukawa term
presented below when $R$ is small, we also expect this theory to
have its origins in string/M-theory, at least for small
curvatures. We have shown that the resulting theory could account
for the accelerated expansion of the universe. We also study our
model in the context of what is known as the scalar-tensor theory
\cite{12} where the metric formalism is used, leading  to a
power-law acceleration in the limit of small curvature. The
Newtonian limit of the theory is investigated and shown to be
correctly satisfied.
\section{Friedmann equation in Palatini formalism }
To begin with, a quick review of the Palatini formalism
\cite{10,11} would be in order. Consider the action
\begin{equation}\label{a}
{\cal S}=-\frac{1}{2\kappa}\int d^{4}x\sqrt{-g}{\cal L}(R)+\int
d^{4}x\sqrt{-g}{\cal L}_{M},
\end{equation}
where $\kappa=8\pi G$ and ${\cal L}_{M}$ is the Lagrangian density
for matter.  Variation with respect to $g_{\mu\nu}$ gives
\begin{equation}\label{b}
{\cal L'}R_{\mu\nu}-\frac{1}{2}{\cal L}g_{\mu\nu}=-\kappa
T_{\mu\nu},
\end{equation}
where a prime denotes differentiation with respect to $R$ and
$T_{\mu\nu}$ is the energy-momentum tensor in the matter frame. We
assume that the universe is filled with dust and radiation,
represented by energy densities $\rho_{m}$ and $\rho_{r}$
respectively, thus
\begin{equation}\label{d}
T_{\nu}^{\mu}=\{-\rho_{m}-\rho_{r},p_{r},p_{r},p_{r}\},
\end{equation}
with $T=g^{\mu\nu}T_{\mu\nu}=-\rho_{m}$, on account of
$p_{r}=\rho_{r}/3$. As is well known, in the Palatini formulation
the connection is not associated with $g_{\mu\nu}$ but with
$h_{\mu\nu}\equiv {\cal L'}g_{\mu\nu}$ which is known from varying
the action with respect to $\Gamma^{\lambda}_{\mu\nu}$. The
Christoffel symbols associated with $h_{\mu\nu}$ are given by
\begin{equation}\label{e}
\Gamma^{\lambda}_{\mu\nu}=\left\{^{\lambda}_{\mu\nu}\right\}_{g}+\frac{1}{2{\cal
L'}} \left[2\delta^{\lambda}_{\,\,\,\,(\mu}\partial_{\nu)}{\cal
L'}-g_{\mu\nu}g^{\lambda\sigma}\partial_{\sigma}{\cal L'}\right],
\end{equation}
where the subscript $g$ indicates association with metric
$g_{\mu\nu}$. The Ricci curvature tensor is given by
\begin{equation}\label{f}
R_{\mu\nu}=R_{\mu\nu}(g)-\frac{3}{2}{\cal
L'}^{-2}\nabla_{\mu}{\cal L'}\nabla_{\nu}{\cal L'}+{\cal
L'}^{-1}\nabla_{\mu}\nabla_{\nu}{\cal L'}+\frac{1}{2}{\cal
L'}^{-1}g_{\mu\nu}\nabla_{\sigma}\nabla^{\sigma}{\cal L'},
\end{equation}
with the curvature scalar written as
\begin{equation}\label{g}
R=R(g)+3{\cal L'}^{-1}\nabla_{\mu}\nabla^{\mu}{\cal
L'}-\frac{3}{2}{\cal L'}^{-2}\nabla_{\mu}{\cal
L'}\nabla^{\mu}{\cal L'},
\end{equation}
where $R_{\mu\nu}(g)$ is the Ricci tensor associated with
$g_{\mu\nu}$ and $R=g^{\mu\nu}R_{\mu\nu}$. Contracting (\ref{b}),
one gets
\begin{equation}\label{h}
{\cal L'}R-2{\cal L}=-\kappa T.
\end{equation}
Equations (\ref{f}) and (\ref{g}) now define the Ricci tensor with
respect to $h_{\mu\nu}$. To derive the Modified Friedmann (MF)
equation we follow \cite{11,13} and consider the flat
Robertson-Walker metric for the evolution of the cosmos
\begin{equation}\label{i}
ds^{2}=-dt^{2}+a(t)^{2}d\textbf{x}^{2}.
\end{equation}
This choice is considered to be consistent with observations
\cite{2}. From equations (\ref{i}) and (\ref{f}) we get the
non-vanishing components of the Ricci tensor
\begin{equation}\label{j}
R_{00}=3\frac{\ddot{a}}{a}-\frac{3}{2}{\cal
L'}^{-2}(\partial_{0}{\cal L'})^{2} +\frac{3}{2}{\cal
L'}^{-1}\nabla_{0}\nabla_{0}{\cal L'},
\end{equation}
\begin{equation}\label{k}
R_{ij}=-\left(a\ddot{a}+2\dot{a}^{2}+{\cal
L'}^{-1}\Gamma^{0}_{ij}\partial_{0}{\cal L'}+\frac{a^{2}}{2}{\cal
L'}^{-1}\nabla_{0}\nabla_{0}{\cal L'}\right)\delta_{ij}.
\end{equation}
Substituting equations (\ref{j}) and (\ref{k}) into field
equations (\ref{b}), we find
\begin{equation}\label{l}
6H^{2}+3H{\cal L'}^{-1}\partial_{0}{\cal L'}+\frac{3}{2}{\cal
L'}^{-2}(\partial_{0}{\cal L'})^{2}=\frac{\kappa(\rho+3p)-{\cal
L}}{{\cal L'}},
\end{equation}
where $ H \equiv\frac{\dot{a}}{a}$ is the Hubble parameter with
$\rho$ and $p$ being the total energy density and total pressure
respectively. If we solve $R$ in terms of $T$ from equation
(\ref{h}) and substitute it into the expressions for ${\cal L'}$
and $\partial_{0}{\cal L'}$ in the above equation, we will get the
MF equation.
\section{Yukawa term}
To continue, let us consider an action of the form
\begin{equation}\label{m}
{\cal L}(R)= R-\frac{\beta}{R}\exp(-\alpha R),
\end{equation}
where $\alpha,\beta>0$ and, as can clearly be seen, the Yukawa
term dominates at small curvature. This feature is important
since, as well shall see, it helps to explain the accelerated
expansion together with the assumption that $R<0$. The contracted
field equation (\ref{h}) now reads
\begin{equation}\label{n}
f(R)\equiv\frac{-R}{\beta}+\alpha\exp(-\alpha
R)+\frac{3}{R}\exp(-\alpha R)= -\kappa T/\beta=\frac{\kappa
\rho_{m}}{\beta},
\end{equation}
where $f(R)$ is a monotonically decreasing function and we have
that $\lim _{R\rightarrow-\infty}f(R)\rightarrow +\infty$ and
$\lim _{R\rightarrow 0^-}f(R)\rightarrow -\infty$. Thus $R$ is
uniquely determined for any value of $\kappa \rho_{m}/\beta\equiv
x$ through equation (\ref{n}). From the conservation equation
\begin{equation}
\dot{\rho}_{m}+3H\rho_{m}=0,
\end{equation}
we find
\begin{equation}\label{o}
\partial_{0}{\cal L'}=\frac{3\beta\left(\alpha^{2}/R+2\alpha/R^{2}+2/R^{3}\right)}
{-\left[\alpha^{2}+3(\alpha/R +1/R^{2})+\exp(\alpha
R)/\beta\right]} \left(\frac{\kappa\rho_{m}}{\beta}\right)H.
\end{equation}
Let us now define $\frac{\partial_{0}{\cal L'}}{{\cal L'}}\equiv
F(x,\alpha,\beta)H$ such that in the limit $\beta\longrightarrow
0$, $F(x,\alpha,\beta)=0$, since from (\ref{o}) one can see that
$\partial_{0}{\cal L'}$ vanishes in this limit. Noting that
$R=R(x)$ and substituting this into equation (\ref{l}) we get the
MF equation
\begin{equation}\label{p}
H^{2}=\frac{\kappa\rho_{m}+2\kappa\rho_{r}-\beta\left[R/\beta-\exp(-\alpha
R)/R\right]}{\left[1+\frac{\alpha\beta\exp(-\alpha R)}{R}+
\frac{\beta\exp(-\alpha
R)}{R^{2}}\right]\left[6+3F(x,\alpha,\beta)(1+1/2F(x,\alpha,\beta))\right]}.
\end{equation}
It can be seen from equations (\ref{n}), (\ref{o}) and (\ref{p})
that when $\beta\longrightarrow 0$ or equivalently
$F(x,\alpha,\beta)=0$, the MF equation will reduce  to the
standard Friedmann equation
\begin{equation}\label{q}
H^{2}=\frac{\kappa(\rho_{m}+\rho_{r})}{3}.
\end{equation}
Let us now consider the evolution of the universe in vacuum, {\it
i.e.} $\rho_{m}=0$ and $\rho_{r}=0 $. If one defines the parameter
$k>1$ according to the following relation
\begin{equation}\label{r} R_{0}=\frac{-1}{\alpha}\ln k,
\end{equation}
where the subscript zero  represents association with vacuum and
substitutes this into the vacuum field equation $f(R)=0$, one gets
\begin{equation}
\alpha=\pm\frac{\ln k}{\sqrt{k\beta(3-\ln k)}}.
\end{equation}
Assuming $\alpha>0$, we find
\begin{equation}
R_{0}=-\sqrt{k\beta(3-\ln k)}.
\end{equation}
Substituting this into the the vacuum MF equation and setting
$x=0$ results in
\begin{equation}\label{s}
H^{2}_{0}=\frac{\sqrt{\beta}}{12}\sqrt{k(3-\ln k)}.
\end{equation}
Note that equation (\ref{p}) reduces to the above equation in
vacuum. This equation tells us that when $\sqrt{\beta}\sim
H^{2}_{0}\sim(10^{-33}eV)^{2}$ and $ k<e^{3}$, modified gravity
with a Yukawa term points to an accelerated evolution of the
universe. It is worth noting that the role of $\sqrt{\beta}$ is
similar to a cosmological constant in this theory. If the density
of dust cannot be ignored, {\it i.e.} $\kappa\rho_{m}/\beta\gg 1$,
then this will be equivalent to $\beta$ being as small as the dust
energy, so that $F(x,\alpha,\beta)\sim0$ and from equation
(\ref{n}) we obtain $R\approx -\kappa\rho_{m}$. Therefore, in the
Big Bang Nucleosynthesis (BBN) epoch, the MF equation reduces to
the standard Friedmann equation. Consequently, the Yukawa term
would no longer be effective there. However, for the present epoch
the condition $\kappa\rho_{m}/\beta\gg 1$ breaks down and the
Yukawa term starts to dominate.
\section{Scalar-tensor formulation}
In this section we study the scalar-tensor formulation of our
model along the lines introduced in \cite{9}. In this theory, the
Palatini form of the action is shown to be equivalent to a
scalar-tensor type theory from which the the scalar field kinetic
energy is absent. This is achieved by introducing a conformal
transformation in which the conformal factor is taken as an
auxiliary scalar field. We show that the resulting field equations
in the small curvature regime predict a power-law type
acceleration for the universe. Let us then first give a brief
review of the formalism discussed in \cite{12}.

Introducing the auxiliary scalar field $\phi$, one may write the
action (\ref{a}) as \cite{12,14}
\begin{equation}\label{eqq2}
{\cal S}[g_{\mu\nu},\phi,\psi_{m}]=-\frac{1}{2\kappa}\int
d^{4}x\sqrt{-g}[{\cal L}(\phi)+(R-\phi){\cal L}'(\phi)]+ {\cal
S}_{m}[g_{\mu\nu},\psi_{m}],
\end{equation}
where $\psi_{m}$ represents the matter field. The equation of
motion resulting from this action for the scalar field is
$\phi=R$, as long as ${\cal L}''(\phi)\neq0$, making action
(\ref{eqq2}) classically equivalent to the action (\ref{a}). This
action is sometimes referred to as the Jordan (physical) frame
action. Under the conformal transformation
\begin{equation}\label{eqq3}
\tilde{g}_{\mu\nu}=e^{\sqrt{\frac{2\kappa}{3}}\Phi}g_{\mu\nu},
\end{equation}
with
\begin{equation}\label{eqq3}
\Phi=\sqrt{\frac{3}{2\kappa}}\ln\cal{L'(\phi)},
\end{equation}
action (\ref{eqq2}) can be written in the Einstein frame as
\cite{12,14}
\begin{equation}\label{t}
\tilde{{\cal S }}[\tilde{g}_{\mu\nu},\Phi,\psi_{m}]=\int
d^{4}x\sqrt{-\tilde{g}}\left[-\frac{\tilde{R}}{2\kappa}-
\frac{1}{2}\left(\tilde{\nabla}\Phi\right)^2-V(\Phi)\right]+{\cal
S}_{m}\left[\exp\left(-\sqrt{2\kappa/3}\Phi\right)\tilde{g}_{\mu\nu},\psi_{m}\right].
\end{equation}
Here,  $\tilde{g}_{\mu\nu}$ is the metric in the Einstein frame
\cite{12} and $\tilde{R}$ is the scalar curvature associated with
$\tilde{g}_{\mu\nu}$. It may be shown that the potential is given
by
\begin{equation}\label{eq1}
V(\phi)=\frac{\phi \cal{L}'(\phi)-\cal{L}(\phi)}{2\kappa
{\cal{L}}'^2(\phi)}.
\end{equation}
This is the standard form of the scalar-tensor type theories
mentioned above. It is worth noting that since we are working with
metric connections here, the kinetic term in  action (\ref{t}) is
present, contrary to the case when one works within the framework
of the Palatini formalism.

Lets us now concentrate on the action represented by (\ref{m}) for
which the potential can be written as
\begin{equation}\label{z}
V(\phi)=\frac{\alpha\beta e^{-\alpha\phi}+\frac{2\beta
e^{-\alpha\phi}}{\phi}}{2\kappa\left(1+\alpha\beta\frac{e^{-\alpha\phi}}{\phi}+\frac{\beta
e^{-\alpha\phi}}{\phi^2}\right)^2}.
\end{equation}
When $\phi$ is small, equation (\ref{eqq3}) is approximated by
$\phi\simeq
\sqrt{\beta}\exp\left(-\frac{1}{2}\sqrt{\frac{2\kappa}{3}}\Phi\right)$
and in this limit  potential (\ref{z}) takes the following form
\begin{equation}\label{eq2}
V(\Phi)\simeq\frac{\phi^{3}}{\kappa\beta}\simeq\frac{1}{\kappa}\sqrt{\beta}
\exp\left(-\sqrt{\frac{3\kappa}{2}}\Phi\right).
\end{equation}
As will be shown below, this potential predicts a universe
evolving  with a power-law acceleration. Typical plots of
$V(\phi)$ are given in figure 1. For $\alpha^{2}\beta=0.1$, the
potential has a maximum at $\phi\simeq0.47$. Therefore, if the
universe starts from the maximum, it either asymptotically evolves
to a de Sitter solution, or undergoes a power-law acceleration.
The minimum of the potential occurs at $\phi\simeq-0.67$ and thus
if the universe starts at this point, it is in an anti-de Sitter
phase where the curvature is negative. Therefore, when the
absolute value of $\phi$ becomes small, the universe evolves to a
power-law acceleration at late-times and at
$\phi\longrightarrow-3.98$, $V(\phi)$ is unbounded from above. For
$\alpha^{2}\beta=0.6$ the behavior of the potential for
$\phi\geq0$ is qualitatively the same as in the previous case. In
the region $\phi<0$, when $\phi\longrightarrow-1.99$ the potential
becomes unbounded from below and for $\phi\simeq-2.1$ it has a
maximum.

To proceed further, let us consider evolution of the scale factor
with time. From action (\ref{t}) one gets
\begin{equation}3\tilde{H}^{2}=\kappa(\rho_{\Phi}+\tilde{\rho}_{m}),\label{eq27}
\end{equation}
and
\begin{equation}\Phi''+3\tilde{H}\Phi'+\frac{dV(\Phi)}{d\Phi}-
\frac{(1-3w)}{\sqrt{6}}\tilde{\rho}_{m}=0,\label{eq28}
\end{equation}
where a prime denotes $d/d\tilde{t}$, and
\begin{equation}
\rho_{\Phi}=\frac{1}{2}\Phi'^2+V(\Phi). \label{eq29}
\end{equation}
These are cosmological equations of motion in the Einstein frame
where $w$ is the usual equation of state parameter and
$\tilde{\rho}_{m}$ is the matter density with $\tilde{H}$ being
the Hubble parameter \cite{8,14}. We must now solve the system of
equations (\ref{eq27}) and (\ref{eq28}). We first consider the
case where $\tilde{\rho}_{m}=0$. When the potential is given by
(\ref{eq2}) and $\phi=R$ is small, a solution is given by
\begin{equation}\label{eq3}
\tilde{a}(\tilde{t})\propto\tilde{t}^{4/3},
\end{equation}
and
\begin{equation}\label{eq4}
\Phi\propto-\frac{4}{3}\ln\tilde{t}.
\end{equation}
Here, $\tilde{t}$ is the time coordinate in the Einstein frame,
which is related to the coordinate $t$ in the Jordan frame by
$e^{-\frac{1}{2}\sqrt{\frac{2\kappa}{3}}\Phi}d\tilde{t}=dt$. As a
result
\begin{equation}\label{eq5}
3\tilde{t}^{1/3}=t,
\end{equation}
The power-law acceleration also occurs in the physical (Jordan)
frame
\begin{equation}\label{eq6}
a(t)=e^{-\frac{1}{2}\sqrt{\frac{2\kappa}{3}}\Phi}\tilde{a}\propto
t^{2}
\end{equation}
which is consistent with the result in \cite{8}. Hence, at small
curvatures, the cosmic acceleration is predicted by the
Yukawa-like term.

\begin{figure}
\begin{center}
\epsfig{figure=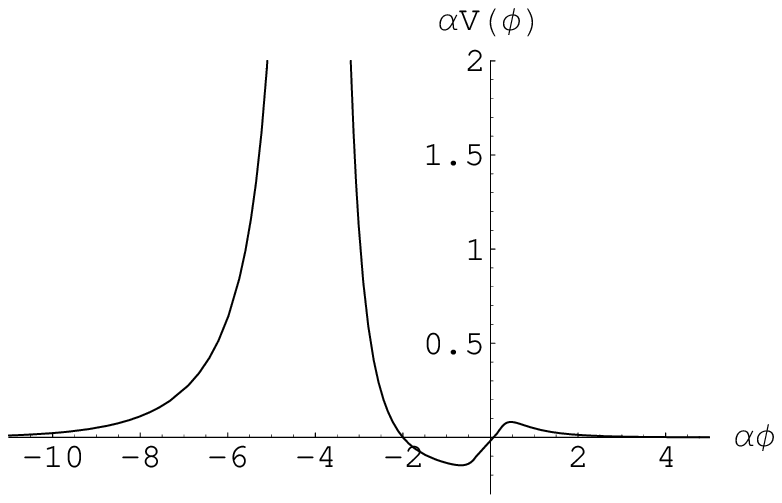,width=6cm}\hspace{5mm}
\epsfig{figure=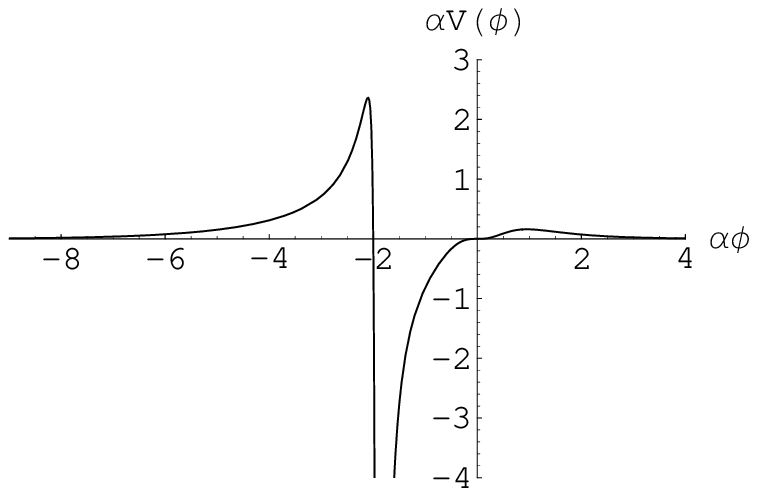,width=6cm}
\end{center}
\caption{\footnotesize The scalar potential given by equation
(\ref{z}) for $\alpha^2\beta=0.1$, left and $\alpha^2\beta=0.6$,
right.}
\end{figure}

Although there is no need to introduce dark energy in modified
gravity, a better understanding of the equation of state in this
theory is afforded by following \cite{linder} and define the
effective Equation of State (EOS). Written in this form, it has
the advantage that any parameterizations done within the context
of EOS  for dark energy can be compared with that of modified
gravity. Let us then start from equation (\ref{eq6}) and find the
Hubble parameter as a function of the redshift $z$ as
\begin{eqnarray}
H(z)=2H_0(1+z)^{1/2}, \label{eq66}
\end{eqnarray}
where $a_0/a=1+z$ with $a_0$ and $H_0$ being the values of the
parameter at the present epoch. It is worth noting that equation
(\ref{eq66}) is the same as that derived from the standard
Friedmann equation with $w=-2/3$. Now, we may write the Friedmann
equation in a formal fashion which would encapsulate any
modification to the standard Friedmann equation in the last term
regardless of its nature \cite{linder}, that is
\begin{eqnarray}
H^2/H_0^2=\Omega_m(1+z)^3+\delta H^2/H_0^2, \label{eq67}
\end{eqnarray}
where
$\Omega_m=\rho/\rho_{0c},\hspace{2mm}\rho_{0c}=3H_0^2/\kappa$.
Also, defining the effective EOS, denoted by $w_{eff}(z)$, as
\begin{eqnarray}
w_{eff}(z)=-1+\frac{1}{3}\frac{d\ln\delta
H^2}{d\ln(1+z)},\label{eq68}
\end{eqnarray}
we can calculate $w_{eff}(z)$ using equations (\ref{eq66}),
(\ref{eq67}) and (\ref{eq68}) with the result
\begin{eqnarray}
w_{eff}(z)=-1+\frac{1}{3}\frac{\left[4-3(1+z)^2\Omega_m\right](1+z)}{4(1+z)-\Omega_m(1+z)^3}.\label{eq69}
\end{eqnarray}
Figure 2 shows variations of $w_{eff}(z)$ for various values of
$\Omega_m$. Equation (\ref{eq69}) shows that
$-1<w_{eff}(z)\le-2/3$, which is  consistent with the
observational data when the limits imposed by appropriate
constraints are taken into account. It is therefore relevant at
this point to have a brief look at these constraints and see how
our results can be interpreted. Before doing so however, it should
be mentioned that the model presented here is not expected to
produce an effective phantom behavior, at least for small
curvatures, since for small $R$ it behaves like $R^{-1}$ theories
for which it is impossible to have phantom behavior, although an
effective quintessence behavior is feasible \cite{abdalla}.

The present observational data \cite{pad1} suggest that our
universe is dominated by a mysterious form of dark energy. As a
result, the universe expansion is undergoing a period of
acceleration. In terms of the constant EOS parametrization,
observational data provided by SDSS indicate that this constant is
close to $-1$. In other words, the accelerating universe could be
either caused by the cosmological constant $w=-1$, or
quintessence, $-1<w<-1/3$ or phantom era, $w<-1$. These
constraints have been verified by the Supernovae, WMAP and cluster
abundance  observations \cite{pad2}. The above results, derived
within the framework of a scalar-tensor type model, seems to favor
the quintessence which predicts values for $w$ in the range given
above. However, in the Palatini formulation presented in section
3, we have obtained the cosmological constant represented by
equation (\ref{s}), which seems to be more in line with the
cosmological constant type models.

\begin{figure}
\begin{center}
\epsfig{figure=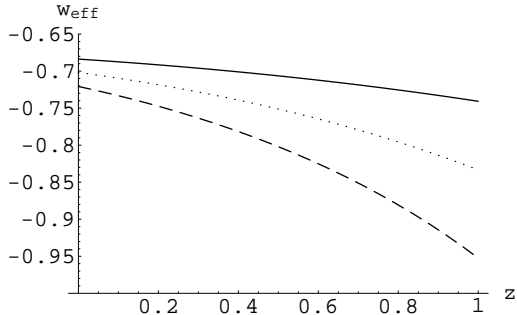,width=7cm}
\end{center}
\caption{\footnotesize Variation of $w_{eff}(z)$ for
$\Omega_m=0.1$ solid line, $\Omega_m=0.2$ dotted line and
$\Omega_m=0.3$, dashed line. }
\end{figure}
\section{Newtonian limit}
The Newtonian limit of theories that predict cosmic acceleration,
that is in the limit of weak field approximation, is naturally of
interest and should be examined since such limits are expected to
be compatible with the present cosmological data. It is known that
the criteria for the correct Newtonian limit of ${\cal L}(R)$
theories is provided by the Dick's condition for fourth order
theories \cite{15}, that is ${\cal L''}(R)|_{R=R_0}=0 $. The same
condition has also been obtained in the Palatini formalism when
the equations of motion are of the second order \cite{dom}. With
this criteria in mind, inspection of our Lagrangian, equation
(\ref{m}), suggests that it does not satisfy Dick's condition.
However, this should not be considered as a setback since with a
small modification which does not affect our general results, we
should be able to obtain the correct Newtonian limit.

Let us then modify Lagrangian (\ref{m}) and write it as
\begin{equation}
{\cal L}(R)=R-\frac{k\alpha^3}{(\ln k)^3}\beta
R^2-\beta\frac{e^{-\alpha R}}{R}. \label{eqq30}
\end{equation}
Now, substitution of the above Lagrangian in equation (\ref{h})
leads to $R_{0}=\frac{-1}{\alpha}\ln k$. Thus for $k=e^{2}$ we
have that ${\cal L''}(R_{0})=0 $. This means that for the choice
of $k$ made above,  we recover the correct Newtonian limit. The
interesting point is that if  one applies the Palatini formalism
to the above Lagrangian, the second term does not contribute in
general and Lagrangian (\ref{eqq30}) behaves exactly the same as
our original Lagrangian does so that the predictions for a
universe with accelerated expansion are unaltered.
\section{Conclusions}
In this paper we have discussed the Palatini formulation of
modified gravity with a Yukawa like term. This term may be used to
explain the current exponential accelerated expansion and  reduces
to the standard Friedmann equation in vacuum.  The scalar-tensor
formulation of our model was discussed and shown  to predict a
power-law acceleration at late times. We have shown that in both
formulations the accelerated expansion of the cosmic may be
accounted for. However, we expect the universe to pass through a
matter dominated era before the accelerated expansion phase is
reached \cite{capo}.

The Newtonian limit of this theory was also examined. It was shown
that in the Palatini formulation our modified gravity action with
a Yukawa term is equivalent to Lagrangian (\ref{eqq30}), with the
latter satisfying Dick's condition for the correct Newtonian
limit.

\end{document}